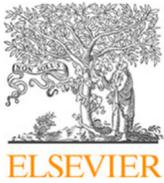
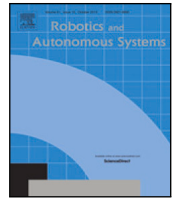

# A SysML-based language for evaluating the integrity of simulation and physical embodiments of Cyber–Physical systems

Wojciech Dudek [a],[*], Narcis Miguel [b], Tomasz Winiarski [a]

[a] *Warsaw University of Technology, Institute of Control and Computation Engineering, Poland, Nowowiejska 15/19, Warsaw, 00-665, Poland*
[b] *PAL Robotics, C/Pujades 77-79, Barcelona, 7-7, Spain*

A B S T R A C T

Evaluating early design concepts is crucial as it impacts quality and cost. This process is often hindered by vague and uncertain design information. This article introduces the SysML-based Simulated–Physical Systems Modelling Language (SPSysML). It is a Domain-Specification Language for evaluating component reusability in Cyber–Physical Systems incorporating Digital Twins and other simulated parts. The proposed factors assess the design quantitatively. SPSysML uses a requirement-based system structuring method to couple simulated and physical parts with requirements. SPSysML-based systems incorporate DTs that perceive exogenous actions in the simulated world.

SPSysML validation is survey- and application-based. First, we develop a robotic system for an assisted living project. We propose an SPSysML application procedure called SPSysAP that manages the considered system development by evaluating the system designs with the proposed quantitative factors. As a result of the SPSysML application, we observed an integrity improvement between the simulated and physical parts of the system. Thus, more system components are shared between the simulated and physical setups. The system was deployed on the physical robot and two simulators based on ROS and ROS2. Additionally, we share a questionnaire for SPSysML assessment. The feedback that we already received is published in this article.

## 1. Introduction

### 1.1. Problem statement

Simulation is widely used in state-of-the-art cyber–physical systems (CPS) development procedures. Recent papers refer to Model-in-the-loop (MIL) [1], Software-in-the-loop (SIL) [2], Hardware-in-the-loop (HIL) [3] and Rapid Control Prototyping (RCP) [4] techniques. They are used selectively or are composed in a sequence, e.g. verification and validation steps in a systems development approach called V-model [5]. Each of the techniques requires a simulation of the system parts. Some systems utilise the Digital Twin (DT) [6,7] concept. They employ an accurate simulation of a system part, e.g., swapping malfunctioned parts, energy consumption analysis, technology integration, or real-time monitoring. Making robot control systems trustworthy is one of the biggest challenges [8]. Application of DT-based testing to robot system development is a promising and popular solution to this problem. However, DT integration to robotic systems raises, e.g., synchronisation issues [9] and integrity issues between DT and the real system software. The software integrity issues are tackled in this article.

The system's parts (software and hardware) interact with the real world, while others interact with the simulated world. We define the set of the parts used in the real world as the physical embodiment and the set of the parts used in the simulated world as the simulated embodiment. The DT is a part of the simulated embodiment mirroring a part of the physical embodiment. DT concept is used in numerous domains and applications [10]. They are systematically analysed by the authors of [11].

Robots comprise multiple devices and their controllers. In particular, autonomous mobile robots require a complex navigation system featuring multiple closed control loops, e.g. drive controllers, trajectory controllers, and Simultaneous Localisation and Mapping (SLAM). Furthermore, in a complex mobile manipulator system [12], the navigation system is integrated with manipulation control to execute user requests (like object transportation). In some cases, only a fraction of such a complex system must have DTs. In another case, if the robot system has the simulated embodiment only, it is a demonstrator of a future product.

To clear up the taxonomy of systems featuring simulated and physical parts, we introduce the Simulated–Physical System (SPSys)

* Corresponding author.
  *E-mail address:* wojciech.dudek@pw.edu.pl (W. Dudek).

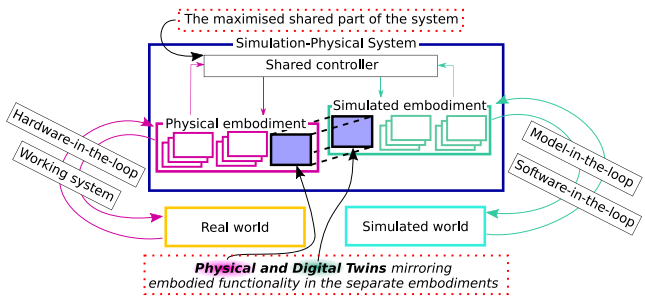

**Fig. 1.** The idea of Simulated–Physical System.

concept (Fig. 1). This kind of system consists of at least one physical/simulated embodiment and a shared controller. If it has both embodiments, the simulated part can be a DT of a physical part. If it has only the physical embodiment, it is a CPS; if it has only the simulated embodiment, it is a simulator.

Developing reliable systems, especially complex ones like robot systems, requires well-designed system architectures [13,14], often defined as Domain-specification language (DSL) and comprehensive unit and integration testing. Some CPS parts can be tested with simulated hardware only; therefore, additional parts that do not comprise the operational system are required (e.g. human simulator in a social robot case). Comprehensive testing is complex in test case specification and time-consuming in test implementation and execution. From this perspective and from issues in robot application in Industry 4.0 [15], software reusability is a key to the fast development of complex, reliable systems.

*1.2. Goal*

We aim to:

- Enhance system requirements traceability for simulated and physical system parts thanks to the introduction of profiled-based requirements of Simulated–Physical Systems.
- Evaluate the level of component re-use between DT and its Physical Twin (PT) and between different system setups, improving the system's reliability and resulting in more accessible and faster testing,
- Model dynamic simulation environment enabling DTs to observe exogenous actions in simulation. The actions can be executed by any agent outside the physical system, e.g., a human moving objects in a service robot environment. Such an action must be implemented in the simulated world to make DT operate in the exact environment as its physical counterpart.

Our goal originates from the conclusion of [16]: *The digital models are mainly used to examine product performance(...). However, how to optimise the use of those models to enhance the design process and design collaboration still needs to be investigated.*

*1.3. Method*

We propose a DSL named Simulated–Physical System Modelling Language (SPSysML) to reach these goals. It is built upon Systems Modelling Language (SysML), the standard and popular language used in the systems engineering domain [17]. SPSysML models the requirements and the SPSys, in particular, SPSys consisting of simulated parts (used either in the system development or in the operational setup as a DT/simulator). Based on SPSysML, we propose a requirement-based SPSys structuring method for enhancing requirement traceability in simulated and physical system parts. To measure the inter-embodiment integrity of the system, we propose quantitative design evaluation factors for SPSys designs. The factors are used during system development to maximise the shared controller, minimise the variety of the system parts and improve simulated–physical integrity.

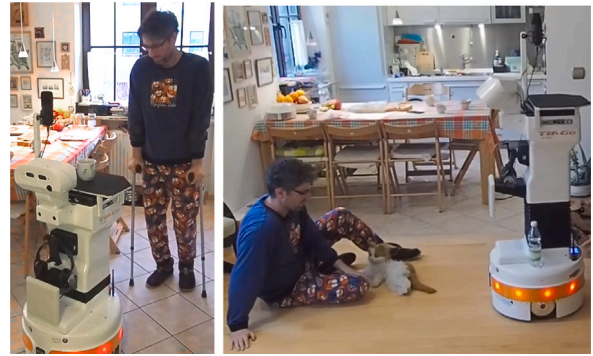

(a) Transport task[1]    (b) Fall assistance task[2]

**Fig. 2.** TIAGo robot in INCARE tasks execution.

In this article, we describe related work concerning mixing simulated–physical setups and meta-models for CPS (Section 2) and state our work's novelty and its result regarding the related works (Section 3). Subsequently, we define SPSys requirements model in Section 4.1, the meta-model in Section 4.2. In Section 5, we analyse the proposed evaluation factors of SPSys design and describe characteristics of a system scored edge case values of the factors. In Section 6, we describe SPSysML implementation in the typical system development steps. Finally, in Section 7.1, we show our method validation in a complex SPSys development for the INCARE project (Fig. 2) and by robotics practitioners assessment 7.2.

**2. Related work**

The key related works are described in Table 1, where their crucial features, including goals, are compared with SPSysML. Other works are briefly described in the following sections.

*2.1. Simulated–physical system development*

In [24], the authors note the robot development process features parallel tasks conducted by heterogeneous R&D teams. The authors suggest system simulation enables rapid prototyping and an iterative development process. They propose the agile Robot Development (aRD) concept facilitating the integration of the hard real-time part and the non-real-time part of the system utilising a Matlab/Simulink toolchain. The sim2real problem is known and mentioned in the recent research results, e.g. in the domain of self-driving cars [27]. However, to our knowledge, none of the existing works specifies a robotic system meta-model with integrity evaluation for mixed simulated–physical setups of CPS.

There are reviews on CPS [28] and DT [16] development. The CPS-related works focus on controlling a physical object and its model in a simulation environment with partially or fully simulated hardware. For example, the authors of [20] use the DT approach to run the digital embodiment in a safe virtual machine and confront the physical and digital embodiments to spot anomalies caused by a cyber-attack. Another work [25] employs simulation-driven machine learning for robots. However, none of the works proposes a system structure allowing exogenous actions execution in the simulated world.

---

[1] https://vimeo.com/670252925
[2] https://vimeo.com/670246589

**Table 1**
Related works comparison with this work.

| System development method | EIV | Formal meta-model | SCP | SE | System structure configurability | DE | E | QE | MDT | Purpose |
| --- | --- | --- | --- | --- | --- | --- | --- | --- | --- | --- |
| Digital Mockup [18] | – | – | ES | – | Single structure | ✓ | – | – | – | Real time (RT) simulation for HIL testing |
| C2PS [19] | – | Own meta-model | Formal and detailed | BS | Single adaptive | ✓ | – | – | N/D | Digital Twin in the cloud |
| [20] | ✓ | Dolev-Yao [21] | Formal and detailed | – | Single structure | ✓ | – | – | ✓ | Cyber-security, physical–digital twin synchronisation |
| DEVSRT [22] | – | DEVS [23] | General | – | Single structure | – | – | – | – | Simulation to embedded continuous development |
| aRD [24] | – | – | General | – | Multiple setups | ✓ | – | – | – | RT & NonRT parts integration |
| HMLF [25] | – | – | ES | – | Single structure | – | – | – | – | Simulation-driven ML |
| RSHPN [26] | – | RSHPN /Petri net | ES | DI | Single structure | ✓ | – | – | – | Control deadlock check, code generation |
| SPSysML | ✓ | SPSysML /SysML | ES/ SPSysAP (Section 6) | SPE | Multiple setups | ✓ | ✓ | ✓ | ✓ | Structure and simulation-physical integrity evaluation |

EIV – Embodiment integrity verification, SCP – System creation procedure, SE – Structure evaluation, DE – supports simulated and physical components, DI – Deadlock identification, E – If uncontrolled agents of the physical environment (link humans) can be modelled in the simulation, QE – If quantitative evaluation factors for system design are proposed, MDT – If a multi Digital Twin system model is presented, ES – Example system as the method application procedure, BS – The best selection from the previously designed, SPE – Simulated–Physical design evaluation, N/D – Possible, however, not defined.

### 2.2. System design evaluation

Early-stage product development challenges systems engineers in designing accurate structures and architectures. Evaluating design concepts is crucial as it mightly influence both the quality and cost of the final product.

Tools and models supporting designers in multiattribute utility analysis are the core of this research area. An example Methodology for the Evaluation of Design Alternatives (MEDA) [29] was presented already in 1991. It is based on the subjective designer's view of the design alternative attributes, e.g. utility. Nowadays, customers play a crucial role in the design process. The authors of [30] tackle the problems that arise in this situation. Priorities are set, and the best design alternative(s) are chosen using the proposed comparison rules. Multi-criteria decision-making is resolved as a constrained multi-attribute optimisation solution [31] or by integrating rough sets in handling vagueness with grey relation analysis [32]. The system's integrity investigated in our work is another system attribute; thus, our evaluation factors values can provide unbiased, objective input to the design evaluation methods and make the design decision aware of the component reuse level between the system's simulated and physical embodiments.

### 2.3. Meta-models for CPS

Engineers use a specific language to plan, conceptualise, and specify the system. Currently, the state-of-the-art approach is a model-based approach utilising a DSL. There are DSLs supporting verification and testing of, e.g. industry 4.0 plants [33], or agent-based computational systems [34]. Unified Modelling Language (UML) is the most known language; however, System Modelling Language (SysML) [17] is proposed by the Object Management Group (OMG) to support the design, analysis and verification/validation of complex systems comprising software and hardware components. Different Model Driven Engineering approaches exist in the robotics domain [35]. For instance, Embodied Agent-Based Cyber–Physical control systems modelling Language (EARL) [36] is SysML specialisation for robotics which allows analysis and specification of the robotic system properties. It is based on the Embodied Agent approach [37]. There are various design analysis tools based on this approach: top-down communication-based system decomposition [38], and Petri net-based design [39] for control deadlocks analysis [26]. It was applied to robotic and multi-modal human–machine interface [40] systems. This article extends the Embodied Agent approach to the simulated and DT-related components domain. We manage this extra complexity with the proposed requirement-based design method and quantitative system design evaluation.

### 2.4. CPS verification methods and SPSysML

MIL [1], SIL [2], and HIL [3] are subsequent testing stages used to verify the system's parts against the requirements. MIL uses the system's parametric and mathematical models to simulate the system's behaviour. Software controllers are parts of the model. In SIL and HIL, the under-verification software parts control the model and the real hardware accordingly. SPSysML is the language used to structure the model and the actual system based on the requirements. Digital Twins designed using SPSysML can be implemented in simulation software like Simulink or other simulators like Gazebo or O3DE. Next, the model can be run against various test cases to verify the parameters or code. SPSysML is used to evaluate the system structure to maximise the system's reusability between model- and hardware-based verification. Additionally, SPSysML lowers the design efforts in tracing the parts' parameters while designing the test cases. Fig. 1 presents the data flow between SPSysML concepts for each MIL, SIL and HIL testing stage.

### 3. Contribution and result validation

None of the models presented in Section 2 specifies SPSys in general, especially including a physical part without equivalent Digital Twin, simulated parts without equivalent PT, or hybrid SPSys including Twins and non-Twin components. The key novel features delivered by SPSysML are:

- Integrity evaluation with quantitative factors based on SysML-based model,
- Guidance for the simulated–physical integrity increase measured by the quantitative factors,
- Requirement-based system composition method for straightforward requirement satisfaction tracing,

- Implementation of system's design evaluation in the CPS development process,
- Reflection of the dynamic environment in simulation to enable Digital Twins to observe exogenous actions,
- A survey-based assessment of Domain-specification Language,
- Validation of a system design evaluation method in robotics by system implementation in ROS and ROS2.

Based on our experience in SPSys development (TIAGo robot [41–43], Velma robot [44], IRP6 robot[3] [45,46]) and the literature analysis, we propose the SPSysML allowing software integrity evaluation between the system embodiments, and further by iterative system design, maximise the between embodiments integrity. Complex system structuring is a broad topic, and solutions are utilised using various methods and strategies, e.g., top-down, middle-out, and bottom-up. [38] describes a binary communication-focused top-down approach for robotic systems. This article proposes a requirement-based bottom-up system structuring method customised for SPSys.

SPSysML comprises a component-based Platform Independent Model (PIM), thus, can be applied to any SPSys. As the system is developed, e.g. with Robot Operating Framework (ROS), it becomes a Platform Specific Model (PSM) that can be launched in the specified setups. PSM implementation utilises platform-specific tools and software libraries. Therefore, to validate our approach, we design, implement, test and deploy a specific SPSys. It uses the TIAGo service robot [47] for the INCARE (Integrated Solution for Innovative Elderly Care) project.[4] The INCARE robot application demonstration videos (simulated/real) are available.[5] We integrate extended voice interface and additional devices to serve the elderly, e.g. in object transportation (Fig. 2(a)) and fall assistance (Fig. 2(b)). The system implementation is component-based and uses frameworks like *ROS* [48,49], *ROS2* [50], *OROCOS* [51], that are a standard open-source robot control frameworks. We use two open-source simulators *Gazebo* [52] (in particular *Gazebo_ros_control* package [53]) and *O3DE* [54] for DT implementation. Besides validation by application, we present results of SPSysML assessment among systems engineering practitioners. To make the SPSysML application easier for the community, we share the SysML profiles, the SPSysML meta-model and the example INCARE system model.[6] All diagrams describing SPSysML and the INCARE system are also shared [55]. This article also presents the SPSysML application procedure that identifies steps of a quantitative evaluation process based on SPSysML.

## 4. Simulated–physical system

Simulated–Physical Systems Modelling Language (SPSysML) is based on SysML and EARL. The following sections describe SPSysML, which defines a stereotype-based meta-model of requirements and system structure. We derive the behaviour models of an agent and its subsystems from EARL. Description of a DSL requires a notation for multiple blocks or instances of a stereotype (e.g. two instances of «*Agent*»). For this purpose, we append the stereotype with *s* (e.g. «*Agents*»). We refer to an instance of a stereotype using the $part-name$ «*stereotype*» symbol.

### 4.1. SPSysML — requirements meta-model

We define a stereotype-based model of requirements (Fig. 3). The stereotypes explicate SysML requirement diagrams used for designing SPSys with SPSysML.

---

[3] Real: https://youtu.be/wJpFcy99Gh0, Simulation: https://youtu.be/BjwcbSdouHw.
[4] http://www.aal-europe.eu/projects/incare/
[5] https://www.robotyka.ia.pw.edu.pl/projects/incare/
[6] https://github.com/RCPRG-ros-pkg/spsysml

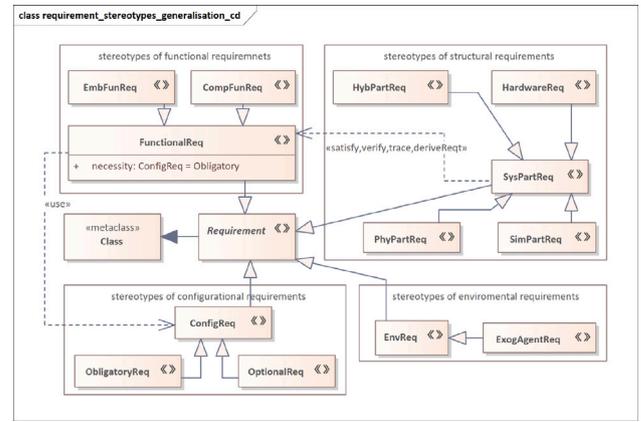

**Fig. 3.** Model of system requirements defined for SPSys.

SPSysML specifies the requirements model as SysML profile. The requirements can be defined based on various premises. In particular, they can derive from user requirements [56]. The SPSysML profile defines structural, functional, configurational and environment requirements. Other requirement types can be added; however, these are used in system setups' specification and requirement-based system composition. Auxiliary sequence diagrams presenting the concept of the system behaviour and use case diagrams are useful for defining requirements. Before SPSys requirements specification, its environment must be analysed. In SPSysML, we distinguish «*ExogAgentReq*» being a part of the environment requirements that specify exogenous agents interacting with the world alongside the system. The system functions given by «*FunctionalReq*» stereotyped requirements are classified into:

- embodied, which requires perceiving or affecting the simulated/physical world — «*EmbFunReq*»,
- computational, which do not interact directly with any of simulated/physical worlds — «*CompFunReq*».

Each «*FunctionalReq*» requirement must be satisfied by a system part; thus «*SysPartReq*» stereotyped requirements must be specified. As shown in Fig. 3, they must satisfy, verify, and derive from — «*FunctionalReq*». In SPSysML, the system structure is based on the requirements of «*SysPartReq*»s stereotype. Each «*SysPartReq*» is classified as elementary «*HardwareReq*» or one of more general «*PhyPartReq*», «*SimPartReq*» or «*HybPartReq*».

- «*PhyPartReq*» determines a part interacting only with the Physical World (even during the system development, e.g. during parallel development of the connected parts), and its simulated embodiment is not required (or cannot be created) during the system development.
- «*SimPartReq*» specifies a part interacting only with the Simulated World (e.g. mock-ups or demonstrators). «*HybPartReq*» interacts with both Worlds (e.g. realised with a pair of DT and PT).
- «*HardwareReq*»s are in a *satisfy* relationship with «*FunctionalReq*»s. This means hardware requirements specified with a «*HardwareReq*» enable the realisation of the functionality specified with the given «*FunctionalReq*». As the system components are specified, the «*HardwareReq*» must be allocated to them. As a result, tracing a component's hardware and allocating to it «*FunctionalReq*» will be straightforward.

The configurational stereotypes specify if a system part («*SysPartReq*») or a system function («*FunctionalReq*») is obligatory to launch the system in all of its setups. Configurational stereotypes are orthogonal to structural and functional stereotypes; thus, functional and structural requirements may be either «*ObligatoryReq*» or «*OptionalReq*».

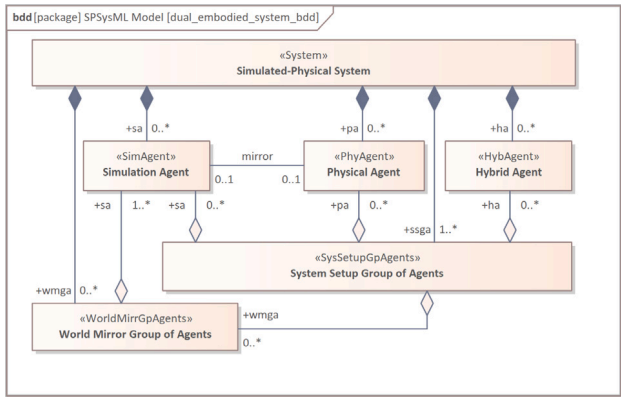

**Fig. 4.** Simulated–Physical System composition.

However, a *«SysPartReq»* realising a *«FunctionalReq»* derives its configurational stereotype from the functional requirement. Each *«Optional-Req,FunctionalReq»* gives the system more setups, and the number of the setups equals $2^n$, where $n$ is the number of *«OptionalReq,FunctionalReq»*. As a result,

- *«ObligatoryReq,FunctionalReq»* requirements define the core system's functions,
- *«OptionalReq,FunctionalReq»* requirements manage the system's versatility by dividing the system into system setups with different functionality,
- *«ObligatoryReq,SysPartReq»* requirements define the system's core components,
- *«OptionalReq,SysPartReq»* requirements define the system's components for different functionality setups.

### 4.2. SPSysML — structure of SPSys

SPSysML derives from EARL [36] version 1.3 [57]. In a SPSys we differentiate three specialisations of the *«Agent»* stereotype defined in EARL (Fig. 4):

- *Physical Agent* (*«PhyAgent»*) – Runs only in the physical embodiment (in particular Agent interacting with or sensing real world),
- *Simulation Agent* (*«SimAgent»*) – Runs only in the simulated embodiment (in particular Agent interacting with or sensing simulated world),
- *Hybrid Agent* (*«HybAgent»*) – Runs in both embodiments and computes without interacting or perceiving any world.

The Groups of Agents called *«WorldMirrGpAgents»*s are Digital Twins of the exogenous agents in the Physical World that are not controlled by the physical embodiment of the system. They execute exogenous actions in the Simulated World. Groups of this type exist if the system works in a dynamic environment (e.g. an environment with human inhabitants). For example, a *«WorldMirrGpAgents»* modifies the Simulated World as humans do in the Physical World.

An Agent is composed of Subsystems of different types as specified in EARL. Types of Subsystems are defined in 4.2.2. We use a Group of Subsystems (*«GpSubsys»*) to gather Subsystems with a specific common properties, and a Group of Agents (*«GpAgents»*) to organise the Agents (composed of Subsystems) cooperating for a defined aim in the system.

#### 4.2.1. Digital Twins in SPSys

Integrity between SPSys embodiments is crucial. Therefore, creating DT of a physical parts is advisable and common. We define the `mirror` relationship between *«PhyAgent»* and *«SimAgent»* to model the relationship between DT and its PT. To enable multi-agent DT for a single *«PhyAgent»* and vice versa, we introduce *«MirrPhyGpAgents»* and *«MirrSimGpAgents»*. They aggregate Agents of different stereotypes (Fig. 5). A Twin Group of Agents is a pair of a Digital Twin and its Physical Twin. Twin Group of Agents comprises *«HybContSubsys»*s shared by the Physical and Digital Twins as shown in Fig. 5.

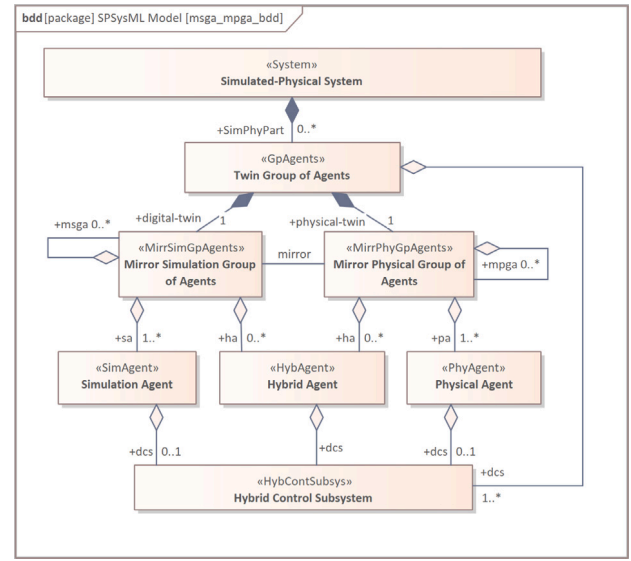

**Fig. 5.** SPSys comprises 0...* Twin Group of Agents, each composed of one Digital Twin and one Physical Twin. The twins are realised with *«MirrSimGpAgents»* and *«MirrPhyGpAgents»* accordingly.

The definition of the `mirror` relationship is as follows:

**Definition 4.1** (*Mirror relationship*). *Two Groups of Agents are said to be in a* mirror *relationship if their input buffers and goals are the same and affect Simulated and Real Worlds congruently. The relationship is an association between Digital and Physical Twins.*

It is worth noting that the same stimuli of the mirroring Groups cause corresponding reactions in specific worlds, in the Simulated World for *Mirror Simulation Group of Agents* and in the Physical World for *Mirror Physical Group of Agents*. The particular reaction results from the requirements of the specific system and does not need to be identical between the Agent Groups that mirror each other. A stimuli for a Group of Agents is the input from other Agents or direct environmental stimuli. Agent's reaction to a stimuli manifests as the Agent state change (including changes in memory or output data).

***Example DT-PT pair***. Considering a pan–tilt head as an example, *«MirrSimGpAgents»* is the simulated head, and *«MirrPhyGpAgents»* is the real one. The stimuli of this DT-PT pair include the light perception from the according world (simulated/physical) and the control for the head joints. The common reaction is data in their output interface. The DT (*«MirrSimGpAgents»*) for a pan–tilt head development is a convenient tool, e.g. for servovision implementation or joint constraints evaluation. SPSys may consist of multiple such pairs, e.g. manipulator and mobile base in a service robot case.

***Constraints and guides for DT-PT pair***. The mirror relationship does not propagate automatically to all the Agents aggregated in the mirroring Groups, so Agents from *«MirrSimGpAgents»* do not mirror all Agents of mirroring *«MirrPhyGpAgents»* and vice versa.

The number of Agents outside DT/PT pairs should be minimised to maximise coverage of simulation-based testing and analysis. Additionally, the complexity and quantity of mirroring Agent Subsystems should be minimised to boost the modularity and integrity of the system. Otherwise, the system could be composed of just two mirroring Agents.

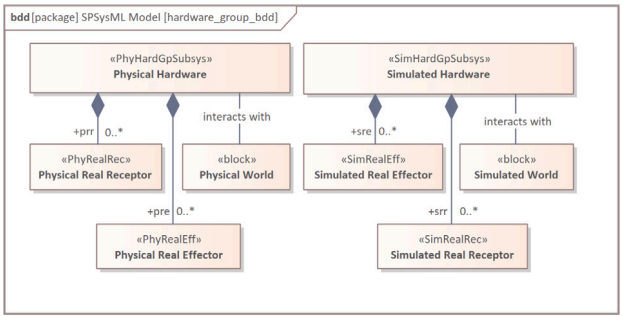

**Fig. 6.** Embodiments of real effectors and receptors.

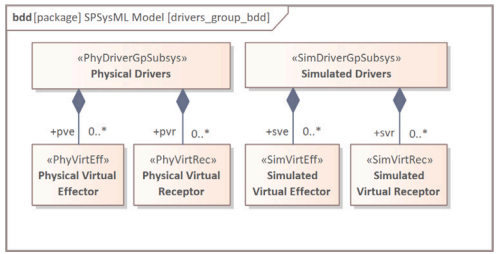

**Fig. 7.** Embodiments of virtual effectors and receptors.

If a computational functionality is required in both embodiments, a *«HybAgent»* should be designed for this purpose.

#### 4.2.2. Subsystem types

The Agents are built with Subsystems of different types. In EARL, the central computational part of an Agent is the Control Subsystem *«ContSubsys»*. An Agent communicates with other Agents using communication buffers of its *«ContSubsys»*. To percept the environment, an Agent uses Real Receptors (*«RealRec»s*), and to aggregate and pre-process stimuli, it uses Virtual Receptors (*«VirtRec»s*). To affect the environment, an Agent uses Real Effectors (*«RealEff»s*). To preprocess *«ContSubsys»* commands to signals for *«RealEff»s*, it uses Virtual Effectors (*«VirtEff»s*). Detailed description and SysML diagrams of Agent structure and behaviour are published in [36].

SPSys interacts with both the simulated and physical world; thus, SPSysML differentiates between the types of the above Subsystems — Simulated, Physical, and Simulated–Physical. To achieve maximum integrity between the simulated and physical embodiments, all *«ContSubsys»s* should be of *«HybContSubsys»* stereotype and constitute the general concept of the shared controller, recall Fig. 1. For iterating design purposes, embodiment-specific Control Subsystems concepts (*«PhyContSubsys»* and *«SimContSubsys»*) may be helpful. Virtual and real effectors/receptors can be Simulated or Physical. In SPSysML we group them in four *«GpSubsys»* that aggregate receptors and effectors for each embodiment — simulated and physical (Figs. 6, 7). Real effectors/receptors are hardware parts, and Virtual effectors/receptors are their drivers.

In complex systems, one *«HybContSubsys»* may communicate with many Simulated/Physical Virtual Receptors and Effectors. In this case, the *«HybContSubsys»* is reused in *«SimAgent»* and *«PhyAgent»*, and both utilise embodiment-specific hardware and drivers. The data flow and communication links for *«HybContSubsys»* interacting with Simulated and Physical Worlds are shown in (Fig. 8).

***Agent composition constraints.*** Each Agent type defined in SPSysML aggregates a particular number of Subsystems of a specific type (Fig. 9). It should be noted that the Basic EARL meta-model defines only one *«ContSubsys»* for an Agent, and inter-agent communication is handled only by the *«ContSubsys»*. In SPSysML, we differentiate between specific types of *«ContSubsys»* for the embodiments; still, an Agent can aggregate just one specialisation of *«ContSubsys»*.

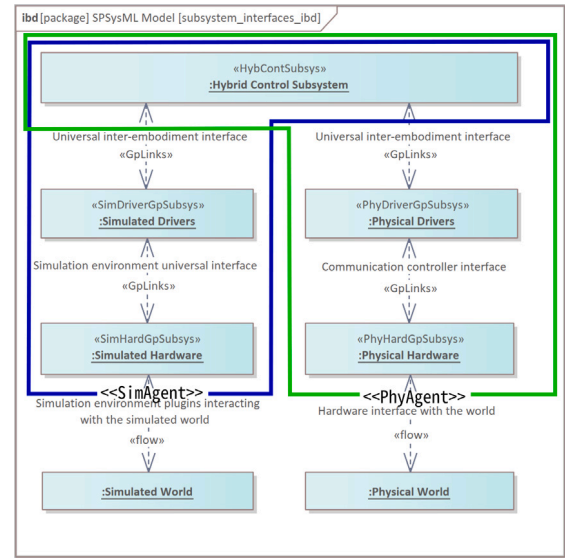

**Fig. 8.** Interfaces between Subsystem Groups on example, where two Agents share *«HybContSubsys»*.

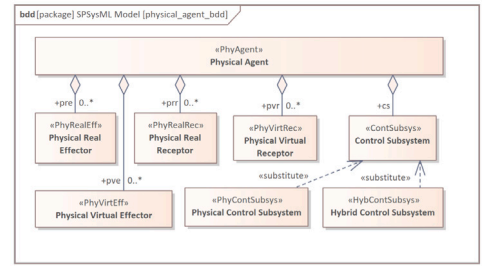

(a) Physical Agent

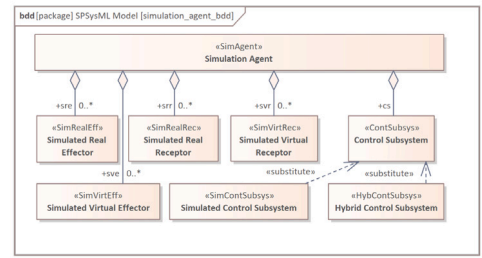

(b) Simulation Agent

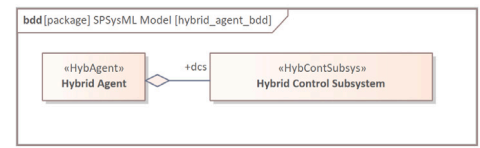

(c) Hybrid Agent

**Fig. 9.** Subsystems aggregated by Agent classes.

***System setups composition.*** SPSys can be deployed in various setups (e.g. testing setups). The setups are a Group of Agents fulfilling the system's functionality in the setup. For the system setup definition, we use *«SysSetupGpAgents»* to specify the Group of Agents working in the setup. The set of Agents composing *«SysSetupGpAgents»* is derived from the system's requirements and test scenarios.

*Example physical embodiment realisation.* Subsystems in a *«PhyHardGpSubsys»* (sensors/actuators) communicate with their drivers via various interfaces (e.g. Linux kernel driver for Inter-Integrated Circuit ($I^2C$) or network interface). *«PhyDriverGpSubsys»*s take data from these interfaces, aggregate it and expose their interface for a *«ContSubsys»* executing the logic of the Agent.

*Example simulated embodiment realisation.* Subsystems in *«SimHardGpSubsys»* expose Application Programming Interface (API) classes, defined in the simulation environment (e.g. gazebo::ModelPlugin for *«SimRealEff»* and gazebo::SensorPlugin for *«SimRealRec»* in Gazebo[7]). Subsystems of *«SimDriverGpSubsys»* connect to these interfaces using the API, aggregate sensor data, and respond *«ContSubsys»*s control.

The connections between the groups are shown in Fig. 8.

## 5. The design evaluation factors

We recognise the following integrity factors classified for design analysis in two scopes:

- System-wide:
  - **Controller integrity factor** (IIF=$\frac{c^{HCS}}{c^{All}}$), where $c^{HCS}$ and $c^{All}$ are the cardinalities of *«HybContSubsys»*s and all system *«ContSubsys»*s accordingly,
  - **Driver generalisation factor** (DGF=$\frac{r_u}{r}$), where $r_u$ is the count of Real Subsystems aggregated in an Agent controlled by a *«HybContSubsys»* and $r$ is the count of all Real Subsystems in the system,
  - **Digital Twin coverage** (DTC=$\frac{a_P^m}{a_{All}^P}$), where $a_P^m$ is the count of *«PhyAgent»* aggregated in a *«MirrPhyGpAgents»* being a PT of a DT (*«MirrSimGpAgents»*), and $a_{All}^P$ is the count of all *«PhyAgent»* in the system,

- DT/PT pair-wide:
  - **Mirror integrity factor** (MIF$_n$ = $\frac{c_n^{HCS}}{c_n^{All}}$), where $n$ is the considered pair of mirroring *«MirrPhyGpAgents»* and *«MirrSimGpAgents»* composing one DT/PT pair, $c_n^{HCS}$ and $c_n^{All}$ are the counts of *«HybContSubsys»*s and all *«ContSubsys»*s accordingly in $n$th DT.

Below, we present the interpretation of the evaluation factors and the correlation between the factor values and the SPSys features. We do it by describing the characteristics of edge case SPSys that scores maximum or minimum values of the design evaluation factors:

- IIF – is a share of the software controller common between the embodiments. Virtual Subsystems are not considered, as their count may be related to the Real Subsystem counts in each embodiment. It is maximised by the reduction of the number of *«SimContSubsys»*s and *«PhyContSubsys»*s in favour of an inter-embodiment *«HybContSubsys»*s, The higher IIF is, the more software components are shared between the simulated and physical embodiments. At maximum (IIF=1), all hardware abstract parts of the system are common.

  - IIF = 0: There are no *«HybContSubsys»*, only *«PhyContSubsys»* or *«SimContSubsys»*. The system's parts in simulated and physical embodiments are disjunctive. Simulation-based testing is not possible. The system's functions in the simulation may be completely different from those in the physical embodiment.

  - IIF = 1: All Control Subsystems are *«HybContSubsys»*, and there are no *«PhyContSubsys»* or *«SimContSubsys»*. This means all hardware abstract parts of the system are common between its embodiments, and the coverage of simulation-based testing is maximised and allows integration testing in simulation.

- MIF$_n$ – is a share of the system parts common between Physical and Digital Twins composing the $n$th twin pair (managing given *«HybPartReq»*). It is similar to IIF but within the scope of $n$th pair of Physical and Digital Twin. Tips for maximisation of the MIF$_n$ factor are:

  - extraction of common functions as *«HybAgent»*s from *«MirrPhyGpAgents»* and *«MirrSimGpAgents»*
  - and/or redesign of interfaces between a *«SimContSubsys»* and *«SimDriverGpSubsys»* and between a *«PhyContSubsys»* and *«PhyDriverGpSubsys»* to emerge a common *«HybContSubsys»* from the *«SimContSubsys»* and *«PhyContSubsys»*,

- DGF – is a share of Real Subsystems (hardware) controlled by *«HybContSubsys»*s (shared controller). It expresses hardware control integrity between the embodiments.

  - DGF = 0: All Hardware parts are controlled by embodiment-specific Control Subsystems. The causes of this depend on a specific case:
    * For Physical Hardware without a DT, it means the interface to hardware is embodiment-specific; thus, extending the system with a DT of the hardware is complicated and would require adding simulation-specific *«SimContSubsys»*.
    * For Physical Hardware with a DT, it means the Hardware Drivers interface of Physical and Digital Twins differ, and the software using the interface differs between the embodiments. This means the system part designed as DT of the Physical hardware is not a proper DT.

  - DGF = 1: All Hardware parts are controlled by embodiment-abstract Control Subsystems; thus, the Agents managing hardware are interchangeable between the embodiments, or future Digital/Physical Twin integration for Physical/Simulated Hardware is straightforward.

- DTC – is a share of hardware and its controllers mirrored with a DT. Its increase boosts coverage of simulation-based testing of hardware controllers and system robustness utilising the DT concept. If DTC = 0, there are no DTs in the system; if DTC = 1, all Physical Hardware parts have DTs.

Based on the factors' values, one can evaluate the system design in terms of:

- Software reusability between the embodiments – based on IIF (for system scope), MIF$_n$ (for $n$th DT),
- Simulation-based testing and failure examination/prediction of the system parts – based on DTC
- Inter-embodiment integrity of hardware controllers and readiness for simulation-based hardware testing – based on DGF.

Maximisation of these factors is not always required, and the optimisation goal can be set at a different point in the factors' space. The goal depends on the specific system requirements. However, the factors' values inform the designer about the inter-embodiment integrity of the system design, so her/his decision is conscious.

---

[7] a popular simulation environment used in the DARPA Robotics Challenge, in July of 2013

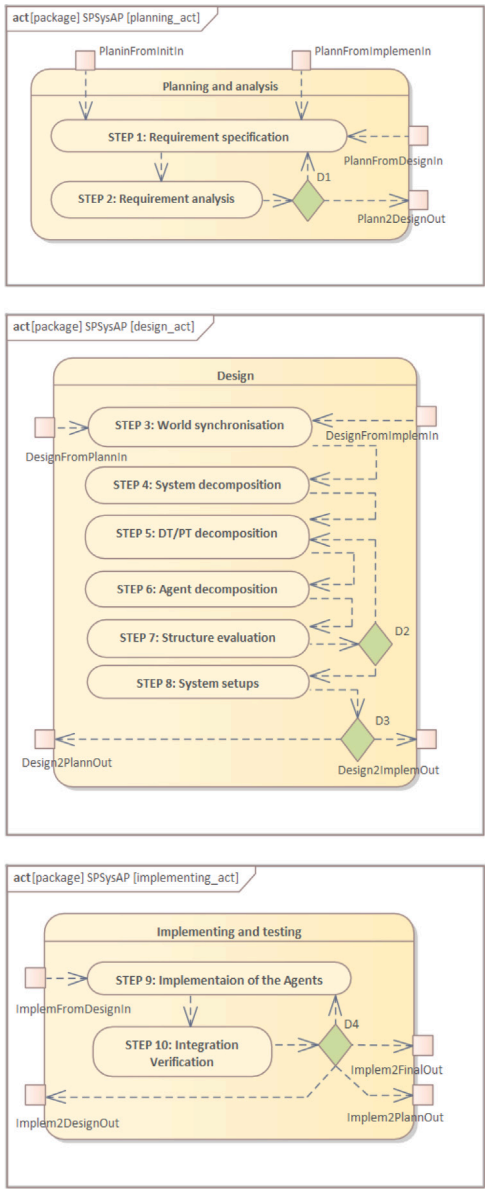

Fig. 10. SPSysML application procedure.

## 6. SPSysML application procedure (SPSysAP)

SPSysML defines system parts and the relationships between them. Various system development procedures can be used for SPSysML-based systems; however, this section proposes a general procedure for integrating SPSysML in the system's development. It enables iterative optimisation of the system's inter-embodiment integrity (Fig. 10). The main activities of the procedure can be executed with different development approaches, traditional, agile, or hybrid. In agile and hybrid approaches, the system is specified and implemented partially in an iterative manner. In the procedure shown in Fig. 10, there are four decision nodes D1, D2, D3, and D4. Implementing SPSysML to exact development procedure requires configuring the logic predicates of these nodes. D1 checks if the requirements of the considered system part (or design iteration) are comprehensive and if the system parts and functions, considering the system's test scenarios, are expressed in the requirement diagrams. D2 checks if the structure evaluation is satisfactory for the considered system part (or design iteration). D3 checks whether any detailed analysis or requirements modification is needed based on the design iteration, and D4 checks if the required *«SysSetupGpAgents»s* are implemented and tested successfully. If needed, the procedure enables complete reiteration from the design stage.

**System planning and analysis**

- **Step 1** (Requirement specification): The requirements are specified using the model defined in Section 4.1. It is recommended that system use cases and typical interactions between system parts be defined on sequence diagrams while specifying the requirements. The system's effect and processes are the centre of gravity for the system definition. Thus, commonly *«FunctionalReq»* show up first. For each of them, an analysis is done if any hardware (*«HardwareReq»*) is needed to satisfy the *«FunctionalReq»*. Each *«HardwareReq»* is analysed if the hardware should work only in simulation setups (*«SimPartReq»*), physical (*«PhyPartReq»*) or both (*«HybPartReq»*). Thus, for instance, hardware can be specified as *«HardwareReq,SimPartReq»*.
- **Step 2** (Requirement analysis): Analysis of the requirements considering the project stakeholders' demands and the system's test scenarios. In case it is needed, modifications to these requirements are made. Plenty of tools and techniques are known for supporting this task [58,59].

**System design**

- **Step 3** (World Synchronisation): This step is not applicable if the system works in a static environment (there are no exogenous actions of any non-system agent executed on the system's environment). In other cases, simulation of external agents (e.g. humans) is required, and it is done by $WorldSync$ *«WorldMirrGpAgents»*. Each *«SimAgent»* composed in *«WorldMirrGpAgents»* manages one *«ExogAgentReq»* specified in Steps 1–2. The key design aspects for the *«SimAgent»*s are the Simulated World model perceived by the SPSys' receptors and the actions affecting this World. To model humans as *«SimAgent»*, we propose a framework named *Human Behaviour in Robotics Research* (HuBeRo) [60,61]. It specifies and implements agents mirroring human behaviours and their physical models in simulation (Fig. 11).
- **Step 4** (System decomposition): The system is decomposed into Agents systematically, based on the parts in the requirements. For each *«PhyPartReq»* and *«SimPartReq»* a $part$ *«PhyAgent»* or $part$ *«SimAgent»* are created. For each *«HybPartReq»* that has only *«ComputationalFunReq»* one $part$ *«HybAgent»* is created. For each other *«HybPartReq»*s pairs of DT/PT are created, thus, these parts are realised with a pair of mirroring $part$ *«MirrPhyGpAgents»* and $part$ *«MirrSimGpAgents»* (Fig. 12).
  Requirements in SysML are connected to the system's parts with «allocate» relationship as shown in our example Figs. 17 and 18. In SPSysML, the «allocation» relationship between *«HardwareReq»* and a system's part sets up the «satisfy» relationship between the system's part and the *«FunctionalReq»* connected to the *«HardwareReq»*. Therefore, the backward analysis from the system's part through «allocation» relationship to *«HardwareReq»* and «satisfy» relationship to *«FunctionalReq»* states the hardware and functional sets of requirements for the system's part. This «allocate» and «satisfy» chain is used for tracing and verification of the dependency between hardware blocks through hardware requirements to functional requirements that are satisfied by the actual hardware blocks.
- **Step 5** (DT/PT decomposition – Mirroring Agent Groups specification): Each mirroring $part$ *«MirrPhyGpAgents»* and $part$ *«MirrSimGpAgents»* is iteratively decomposed to more groups to finally reach mirroring groups that each can be realised with a single *«PhyAgent»* or *«SimAgent»*. Each result of the decomposition iteration represents a layer of the initial Group. The $mirror$ relationship is set between *«MirrSimGpAgents»* and *«MirrPhyGpAgents»* and is specified in each layer of the decomposition. The decomposition scheme is shown in Fig. 12.

- **Step 6** (Agent decomposition): Each *«Agent»* is decomposed into Subsystems. Various approaches can be used depending on the requirements formulation; however, we advise a layered bottom-up one with a definition of each layer interface. First, the hardware assigned to each *«Agent»* is expressed as *«PhyRealRec»*s, *«PhyRealEff»*s, *«SimRealEff»*s and *«SimRealRec»*s (constituting Agent hardware layer). Next, the interfaces of *«PhyVirtRec»*s, *«PhyVirtEff»*s, *«SimVirtEff»*s and *«SimVirtRec»*s are defined based on Section 4.2.2 (constituting Agent driver layer). The target is to develop a *«HybContSubsys»* (constituting Agent control layer) that:

  – manages behaviours of both mirroring Agents (simulated and physical),
  – controls hardware (simulated/physical) using the embodiment-common interface of *«PhyDriverGpSubsys»* and *«SimDriverGpSubsys»*,
  – communicates with other Agents using an inter-agent interface.

  However, a temporary embodiment-specific Control Subsystems (*«SimContSubsys»* and *«PhyContSubsys»*) are helpful during the design process and testing. Mirroring Agents consist of identical control layers; however, the Agents differ in drivers and hardware layers. The driver layer must fill the gap between the hardware and controller layers in the specific embodiments to enable embodiment abstraction for the control layer in mirroring Agents.

- **Step 7** (Structure evaluation): In each design iteration, the structure is evaluated.
  The target of the structure optimisation process is the maximisation of the shared-controller part of the system (maximisation of IIF, MIF, DGF, DTC factors). The share of physical parts mirrored with Digital Twins (DTC) additionally boosts the system's robustness, as DT can replace malfunctioned hardware. The detailed analysis of the proposed factors is described in Section 5. The optimisation target can be defined with the above factors or others that can be defined for a specific system. If the evaluation result is unsatisfactory, the system should be redesigned (starting from **Step 5**) following the factors' maximisation guidelines defined in Section 5.

- **Step 8** (System setups): Based on the *optional* stereotypes in the requirements, all possible system setups emerge. Each setup (*«SysSetupGpAgents»*) is specified with an Internal Block Diagram (example diagrams are shown in Figs. 17, 18). The diagram shows the Agents composing given *«SysSetupGpAgents»*, their communication and interaction with Simulated and Physical Worlds. It should be noted that the system's operational setups are just a starting point for *«SysSetupGpAgents»*s specification. For each testing scenario, a *«SysSetupGpAgents»* should be specified. DTs are broadly used in development optimisation and CPS development safety improvement; therefore, in particular, for testing a `part` *«MirrPhyGpAgents»*, a *«SysSetupGpAgents»* consisting of `part` *«MirrSimGpAgents»* (DT of the *«MirrPhyGpAgents»*) should be defined.

**System implementation and testing**

- **Step 9** (Implementation of the Agents): *«SimAgent»*s and *«HybAgent»*s should be implemented before *«PhyAgent»*s for safety reasons. We want to test the system in simulation before its deployment on the real hardware. Before implementation, subsystems must be translated into a Platform-Specific Model (PSM). For robotic systems utilising ROS/ROS2, we propose MeROS DSL [62]. Implementation should include unit testing of each Subsystem and Agent (e.g. applying Robot Unit Testing methodology [63]).
- **Step 10** (Integration verification): Test scenarios execution for all implemented *«SysSetupGpAgents»*.

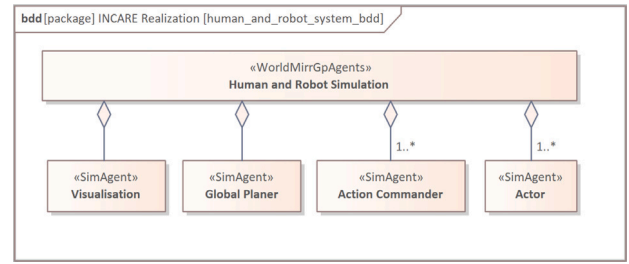

**Fig. 11.** Agent types of HuBeRo framework [60] used in SPSys as *«WorldMirrGpAgents»* that mirrors humans in Simulated World.

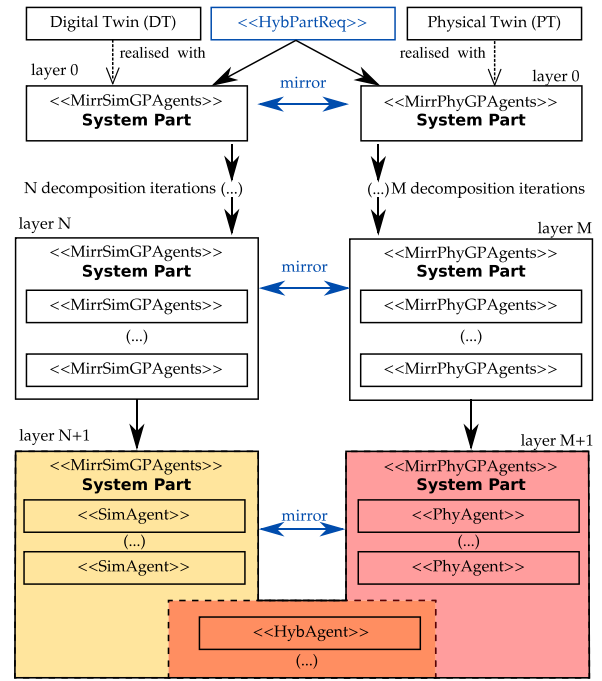

**Fig. 12.** Requirement-based SPSys decomposition, where `mirror` relationship constituting Physical and Digital Twins is obligatory for *«HybPartReq»*. As a result of the decomposition, DT/PT may share *«HybAgent»*s.

## 7. SPSysML validation

First, we describe the SPSysML validation by application. It includes the requirement-based system composition, design evaluation, and applicability in the system development procedure 7.1. Next, we share the results of the SPSysML assessment done among systems engineering practitioners, including a third-party SPSysML user 7.2.

### 7.1. Example system design and analysis

SPSysML was used in developing complex SPSys utilising a service robot for the INCARE project. This is the primary validation of SPSysML. The INCARE system idea and requirements are published in [64]. The framework model developed to manage robot tasks is published in [65,66]. Developing a complex system requires different implementations of its parts in various development phases. In INCARE, we use some commercial products like the TIAGo robot with its control system. We develop and integrate new parts into the system (e.g. human fall detector and TIAGo audio interface extension). We use and modify the TIAGo robot in this project; thus, we specify its hardware and controller as a part of the INCARE system specification. We present the result of the SPSysML application in the following SPSys development tasks.

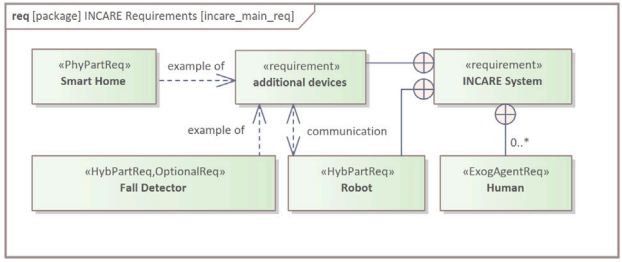

(a) INCARE structural requirements

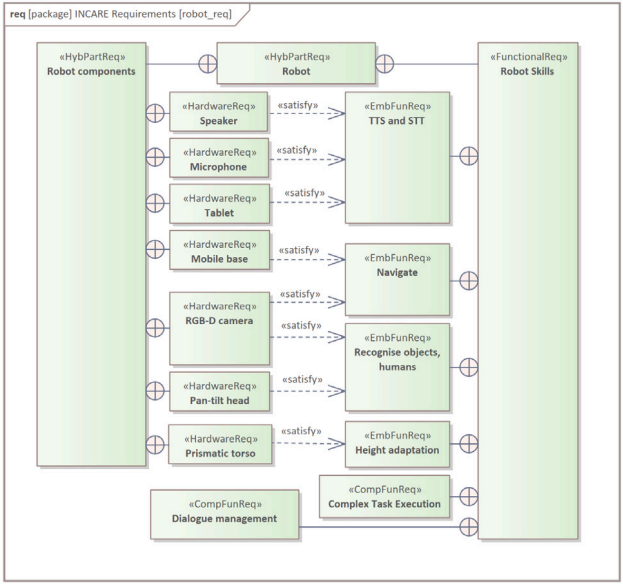

(b) The Robot requirements, where TTS and STT are text-to-speech and speech-to-text functions

**Fig. 13.** Example part of the INCARE requirements.

*Requirement engineering.* In these steps, we use the requirement model defined in SPSysML. We specify the structural and functional requirements based on the general requirements of the INCARE project. The general and the robot-specific requirement diagrams are shown in Fig. 13. The requirements were accepted after some Step 1↔Step 2 iterations. These iterations led, for instance, to the decomposition of the `Communication with humans «EmbFunReq»` to `TTS and STT «EmbFunReq»` and `Dialogue management «CompFunReq»`.

*Exogeneity identification.* The robot in the INCARE project coexists with humans being exogenous agents from its perspective; thus, we utilise `WorldSync «WorldMirrGpAgents»` to manage humans in the Simulated World. Detailed specification and realisation of `WorldSync «WorldMirrGpAgents»` is available in [60].

*System composition.* System structuring resulted in the system being composed of 5 embodiment-specific Agents, one for each system part:

- `TIAGo «SimAgent»`, `TIAGo «PhyAgent»`,
- `FallDetector «SimAgent»`, `FallDetector «PhyAgent»`,
- `SmartHome «PhyAgent»`.

Additionally, there are 3 Hybrid Agents, one for each computational function of the system:

- `ComplexTaskExecution «HybAgent»`,
- `Talker «HybAgent»`,
- `FakeAudio «HybAgent»`.

There are two DT-PT pairs:

- one resulting from the requirement `Robot «HybPartReq»`:
    - `Robot «MirrSimGpAgents»` being a DT composed of `TIAGo «SimAgent»`, `FakeAudio «HybAgent»` and `Talker «HybAgent»`,
    - `Robot «MirrPhyGpAgents»` being PT composed of `TIAGo «PhyAgent»` and `Talker «HybAgent»`.

- one resulting from the requirement `Fall Detector «OptionalReq,HybPartReq»`: `FallDetector «SimAgent»` as DT and `FallDetector «PhyAgent»` as PT.

*Agent decomposition.* To provide an example, we describe the final decomposition of two `TIAGo «SimAgent»` realisations (`O3deTIAGo` (Fig. 14) being the robot simulator implemented in the O3DE simulator and `GazeboTIAGo` (Fig. 15) implemented in Gazebo) and `TIAGo «PhyAgent»` (Fig. 16) which mirrors the previous ones. Each robot hardware component is specified as either *«PhyRealRec»*, *«PhyRealEff»*, *«SimRealRec»*, or *«SimRealEff»*. `O3deTIAGo: TIAGo «SimAgent»` integrates the system with O3DE simulation environment; thus, *«SimRealRec»s* and *«SimRealEff»s* are core O3DE components interacting with the simulated world, called gems. `GazeboTIAGo: TIAGo «SimAgent»` integrates the system with Gazebo simulation environment; thus, *«SimRealRec»s* and *«SimRealEff»s* expose gazebo::SensorPlugin and gazebo::ModelPlugin interfaces accordingly. As the physical robot, we use PAL Robotics' TIAGo; thus, *«PhyRealRec»s*, *«PhyRealEff»s* interfaces are adequate Linux Kernel drivers managing communication with the devices. In `GazeboTIAGo: TIAGo «SimAgent»` case, the Simulated Drivers connect to the gazebo::SensorPlugin and gazebo::ModelPlugin interfaces and expose the robot state information and typical ROS topics/services (e.g. *JointStateInterface* and *EffortJointInterface* for `MobileBaseController «SimVirtEff»`[8] and */scan* ROS topic for `lidar [«SimVirtRec»]`). If Simulated World is the Gazebo environment, *«SimVirtEff»s* and *«SimVirtRec»s* are usually implemented as Gazebo Plugins. *«PhyDriverGpSubsys»* connects to Linux Kernel drivers and, as a whole Group, exposes to *«ContSubsys»* identical interfaces as *«SimDriverGpSubsys»*. The diagrams of the TIAGo robot Agents (simulated and physical) show their common `RobotIf «HybContSubsys»`; however, the physical robot and Gazebo simulator run `ROS1: RobotIf «HybContSubsys»`, and O3DE simulator runs `ROS2: RobotIf «HybContSubsys»`.

*Structure evaluation.* The final structure evaluation resulted with: IIF=1, MIF$_{Robot}$=1, MIF$_{FallDetector}$=1, DGF=1, and DTC=0.67. The result means the structure consists of no *«SimContSubsys»* or *«PhyContSubsys»*, and one *«PhyAgent»* is not mirrored by a DT — `SmartHome «SimAgent»`. The lack of DT for `SmartHome «SimAgent»` results from the requirements (Fig. 13(a)), where Smart Home is not *«HybPartReq»*; thus, as the requirement is not modified, this case is the designer's informed decision. DGF=1 means the considered `SmartHome «SimAgent»` has *«HybContSubsys»*; therefore, its control subsystem can be shared with a DT, if one will be required in the future. One of the previous design iterations resulted with: IIF=$\frac{5}{7}$ = 0.71, MIF$_{Robot}$=1, MIF$_{FallDetector}$=0, DGF=$\frac{20}{22}$ = 0.91, DTC=0.67. In this iteration the `FallDetector «SimAgent»` and `FallDetector «PhyAgent»` use embodiment specific control layer (*«SimContSubsys»* and *«PhyContSubsys»*), because `FallDetector «PhyAgent»` consists *«PhyContSubsys»*. This *«PhyContSubsys»* cannot be shared with DT. To increase MIF$_{FallDetector}$ and DGF the common part of `FallDetector «SimContSubsys»` and `FallDetector «PhyContSubsys»` was extracted. The common part constitutes `FallDetector` «HybContSubsys» in the final design. The `FallDetector` «HybContSubsys» becomes a universal interface between `ComplexTaskExecution` «HybAgent» and the driver layer of `FallDetector` «PhyAgent». It forced decomposition of `FallDetector` «SimAgent» to `FallDetector` «HybContSubsys», `FallDetector` «SimVirtRec» and `FallDetector` «SimRealRec». The latter two simulate the Fall Detector sensing. Thanks to the design evaluation, `FallDetector` «HybContSubsys» is used in both embodiments in the final structure, and this part of `FallDetector` will be tested in simulation because it is common in simulated and physical embodiments of the system.

---

[8] This *«VirtEff»* is based on the *gazebo_ros_control* package: https://classic.gazebosim.org/tutorials?tut=ros_control.

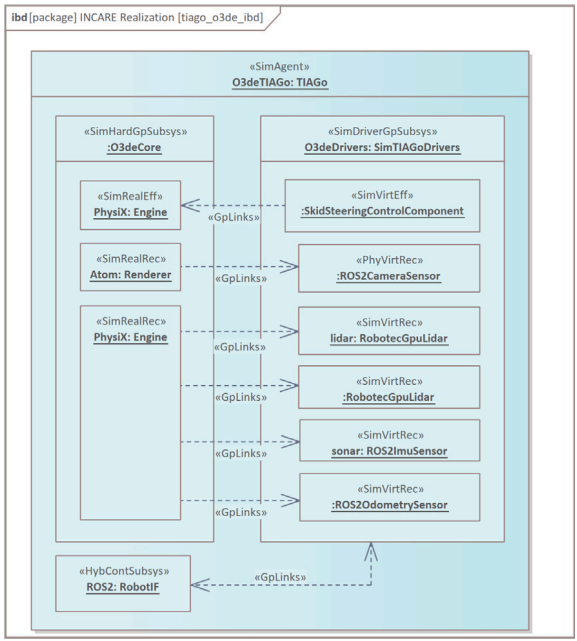

**Fig. 14.** IBD of `O3deTIAGo: TIAGo` «SimAgent».

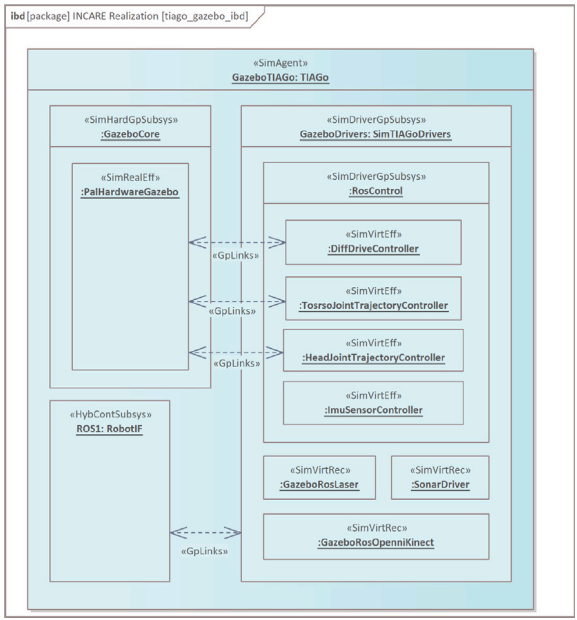

**Fig. 15.** IBD of `GazeboTIAGo: TIAGo` «SimAgent».

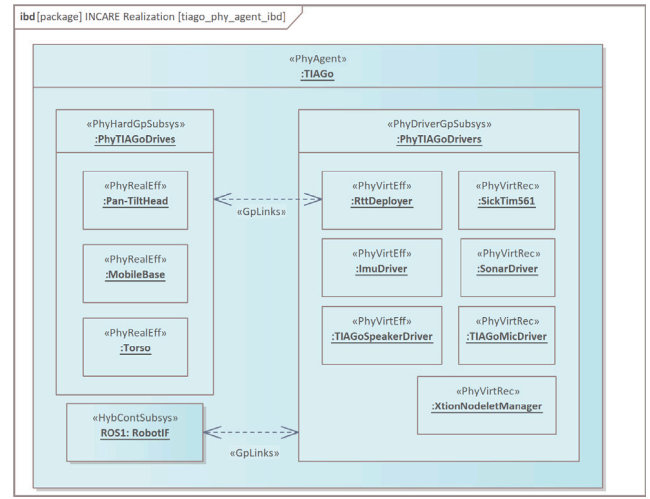

**Fig. 16.** IBD of `TIAGo` «PhyAgent».

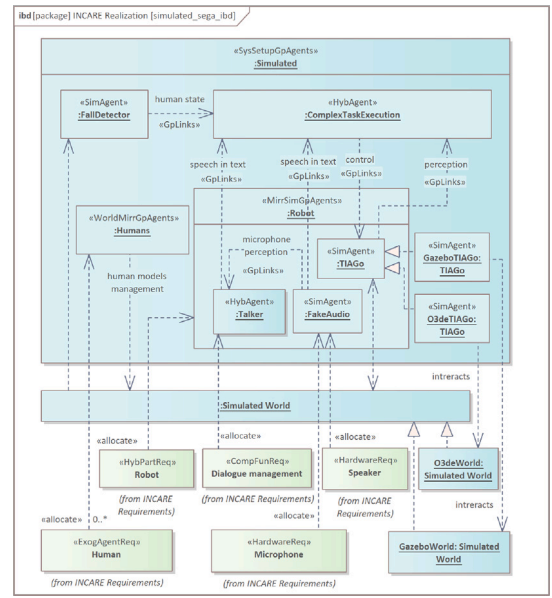

**Fig. 17.** `Simulated` «SysSetupGpAgents» with example requirement allocations.

***Configurability definition.*** Based on the «OptionalReq»s in the requirements we define 6 «SysSetupGpAgents»s composing of:

- `FallDetector` «PhyAgent» or `FallDetector` «SimAgent» or no `FallDetecor`, and
- `Robot` «MirrPhyGpAgents» or `Robot` «MirrSimGpAgents».

We show `Simulated` «SysSetupGpAgents» (Fig. 17) and `Physical` «SysSetupGpAgents» (Fig. 18) Internal Block Diagrams (IBDs) to exemplify «SysSetupGpAgents» specification. `Simulated` «SysSetupGpAgents» presents two realisations of `TIAGo` «SimAgent» — `GazeboTIAGo` and `O3deTIAGo`.

***Implementation.*** In INCARE the `Robot` «PhyAgent» and `Robot` «SimAgent» are realised with TIAGo robot and its simulations. PAL Robotics mostly implemented these. We mapped the robot hardware and software to our design (as shown in Figs. 14–16). `RosControl` «SimDriverGpSubsys» is a software package[9] implementing various common ROS controllers, and `SimTIAGoDrives` [«SimHardGpSubsys»] is a package[10] implementing Gazebo plugins controlling TIAGo drives.

---

[9] http://wiki.ros.org/ros_control
[10] http://wiki.ros.org/pal_hardware_gazebo

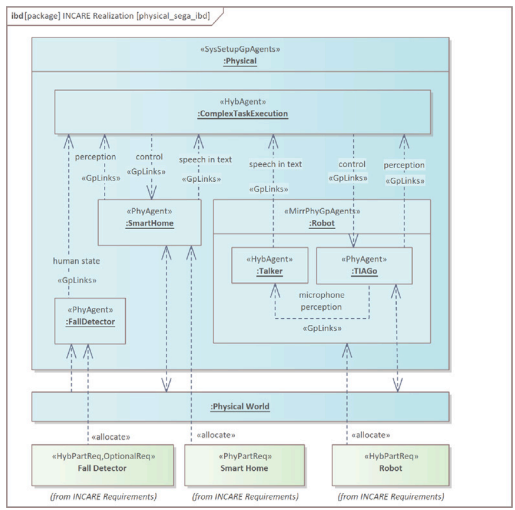

**Fig. 18.** `Physical «SysSetupGpAgents»` with example requirement allocations.

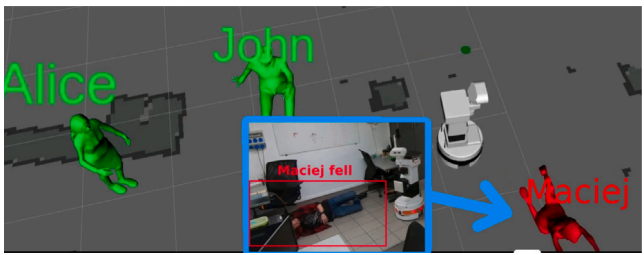

**Fig. 19.** A combined frames from running separate simulation and physical system's embodiments.[11] The robot perceives the simulated humans who are visualised in Rviz.[12]

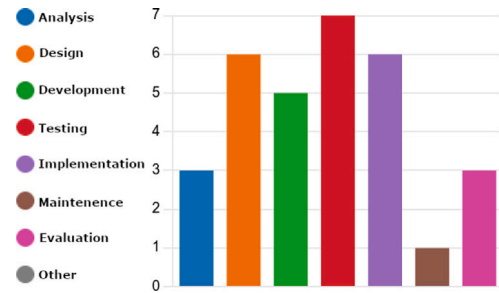

**Fig. 20.** Tasks in which the responders utilise simulation.

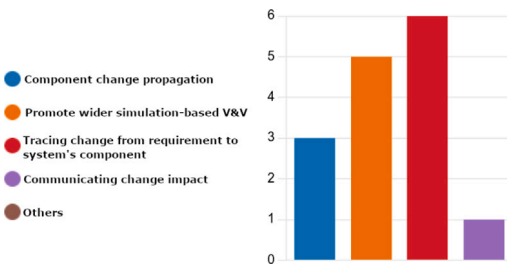

**Fig. 21.** How SPSysML would help you in change analysis of a complex CPS?.

Additionally, we improved the robot's voice communication by integrating a Large Language Model for user request interpretation [43]. We equipped `TIAGo «PhyAgent»` with additional USB Microphones [67] to improve its audio perception. `Complex Task Execution «HybAgent»` is realised with TaskER framework [41,66]. The simulated and physical worlds' synchronisation feature (mirroring humans in the Simulated World) was implemented as `WorldSync «WorldMirrGpAgents»` and the system's simulated and physical embodiments were tested in the according worlds (Fig. 19).

**Validation.** The INCARE system was deployed and validated in an end-user home, and the videos present its performance in the example tasks — transportation (Fig. 2(a))[1] and fall assistance (Fig. 2(b))[2].

### 7.2. SPSysML assessment

Based on the open-source SPSysML documentation, a third-party developer designed a system for quantitative evaluation of autonomous vacuum cleaner navigation algorithm [68]. This design activity scored the best possible evaluation factor values. The designer is sure the requirement-based system structuring helps in comprehensive analysis and would use it again as it simplifies CPS design. He commented on the SPSysML: *SPSysML promotes wider simulation-based V&V of the system's change, and evaluation factors help validate and point out the system's weak parts. However, I am unsure how to implement hybrid agents in ROS, e.g., the same communication topics and structure. Early decomposition is the hardest part of the SPSysML application.*

---

[11] sim: https://vimeo.com/403391725, real: https://vimeo.com/521756050
[12] https://wiki.ros.org/rviz

To evaluate our methodology wider, we asked systems engineering and robotics practitioners to assess SPSysML from various points of view. We designed the questionnaire with the help of a professional poll analyst. The questions regard individual experience (Fig. 20), SPSysML utility (Fig. 21), factor improvement guidance (Fig. 22), drawbacks, difficulties and our methodology coupling possibilities with standard tools like Zachman Framework [69] and V-model [70]. Finally, the responders assessed our methodology against the qualities of great models (Fig. 23). The questionnaire form and all results are published[6]. The example responses include drawbacks:

- *Evaluation factors might be abused as the importance of components is not taken into account*
- *The proposed design evaluation factors are not clearly understandable without a good grasp of SPSysML.*
- *With a large number of requirements that have a complex structure, the diagrams may be unreadable. The solution would be a requirements association table.*

and advantages:

- *SPSysML certainly gives you the opportunity to evaluate the system architecture and make conscious design decisions.*
- *At early stage of system development SPSysML can show some issues regarding system design*
- *Joining simulated and physical models provides numerous benefits. SPSysML allows to clearly depict this in system architecture.*
- *SPSysML is a systematic framework for joining simulation and physical systems. This process can be controlled with SPSysML.*

All responders would apply the requirement-based composition, and eight of nine would apply SPSysML for system specification and design evaluation.

## 8. Discussion

We state that incorporating SPSysML-based quantitative evaluation of system design indicates significant system structure changes resulting from the proposed quantitative factor-based guidelines applications. An example is `FallDetector «SimAgent»` and `FallDetector «PhyAgent»` integration improvement by a common function isolation as *«HybContSubsys»*. Moreover, the verification shows that 33% of the system's Physical Agents do not have DTs, and thanks to the design modification, the whole hardware (physical and simulated) is controlled with common components (DGF=1). This has two main advantages. First, all Digital and Physical Twins pairs share a common control subsystem. Second, if there is a *«PhyAgent»* or *«SimAgent»* without a twin, it can be easily integrated using the already implemented embodiment-abstract interface.

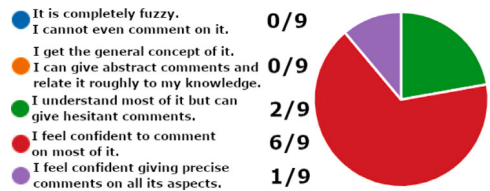

**Fig. 22.** Clarity of the factor-based rules for integrity improvement.

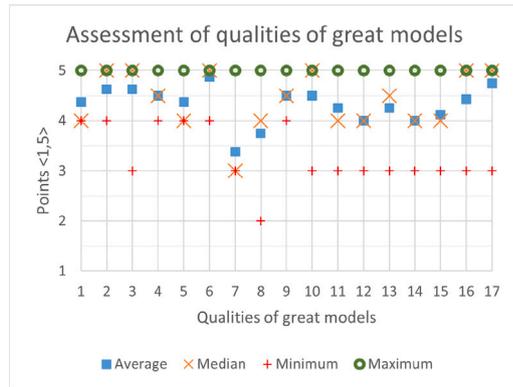

**Fig. 23.** SPSysML assessment against qualities of great models: (1) Linked to Decision Support, (2) Model Credibility, (3) Clear Scope, (4) V&V With Model, (5) Well-Organised, (6) Analysable and Traceable, (7) Data Extrapolation, (8) Complete, (9) Internally Consistent, (10) Verifiable, (11) Validation, (12) Model Fidelity, (13) Elegant, (14) Formed for Optimisation, (15) Avoid Black Box, (16) Availability of Interfaces, (17) Reusable. The qualities are defined in the questionnaire.

We argue that the requirement profile for SPSys enables critical analysis and may result in requirement decomposition and explication. An example is the decomposition of the `Communication with humans «EmbFunReq»` to `TTS and STT «EmbFunReq»` and `Dialogue management «CompFunReq»` during requirement analysis. Thanks to the `Dialogue management «CompFunReq»` separation, the proposed requirement-based procedure for the system composition resulted in the implementation of `Talker «HybAgent»` managing the requirement. As a consequence of the above, the requirement is managed by a component useable in different setups in cooperation with the simulated or physical embodiment. Otherwise, the requirement would be managed separately in simulated and physical embodiments of the robot.

We associate *inter-embodiment integrity* with DT software reusability. The proposed structuring method allows the reveal of the set of universal system components aggregated in the shared controller part of the system. Thus, designing and implementing the DT software components to be used in the actual system increases the software reusability and inter-embodiment integrity. In SPSysML, the "shared controller" is constituted by the set of *«HybAgent»* and *«HybContSubsys»*. The evaluation factors state the DT software reusability (or inter-embodiment integrity) by reflecting the share of "shared controller" in the embodiment-specific set of the system's components. SPSysML is a method for measuring inter-embodiment integrity (or DT software reusability). Therefore, the resolution and accuracy of the measurement depend on the measuring equipment. In this case, the equipment is the system model designed with SPSysML. The measurement is precise if the system comprises elementary components/software classes. However, measuring the evaluation factors is rough if the model consists of several huge and complex components.

The proposed application procedure can be implemented in the complex system development process, and the iterative correction of the system's design using the evaluation factors informs the designer about embodiment-specific components used in the design.

Systems engineering practitioners who reviewed SPSysML confirm the above statements and point out minor difficulties in the SPSysML application.

## 9. Conclusion

Numerous cyber–physical systems include simulated parts such as Digital Twins (DT), demonstrators, or mockups utilised during their development. We postulate the Simulated–Physical System (SPSys) concept to describe this kind of system. SPSys includes physical, simulated and hybrid parts.

SPSys application is wide; in particular, it can work in a dynamic environment and observe exogenous actions of the inhabitants in the environment. Such a situation is problematic because DT must perfectly mirror its PT. To answer the above needs, we propose a Domain-Specific Language named Simulated–Physical System Modelling Language (SPSysML) that defines SPSys taxonomy and the relationships between the types of SPSys parts. We also present the procedure for design evaluation using the SPSysML-based specification of the system under study.

Integrity and reusability in reliable software development is crucial. Therefore, based on SPSysML terms, we propose design evaluation factors and SPSysML application procedure that enable quantitative analysis and optimisation targeted to software re-use maximisation between DT and PT in different system setups. We analyse the evaluation factors and show features of the systems that score edge case values of the factors.

Finally, we validate SPSysML, including the evaluation factors and proposed requirement-based system composition in a complex robot system development. We demonstrate step-by-step results of SPSysML application. The validation shows that the requirement profile for SPSys enables critical analysis and may result in requirement decomposition and explication. The validation also demonstrates the positive influence of the SPSysML-based quantitative evaluation on component reusability and integrity between simulated and physical parts.

The presented validation confirms the features distinguishing our method from the others listed in Table 1, e.g.:

- Quantitative evaluation of integrity between the system's simulated and physical embodiments,
- Modelling simulation, physical and hybrid system execution and testing setups,
- SysML-based specification of systems composed of physical and simulated parts and Digital Twins.
- Enabling Digital Twins to observe exogenous actions in the simulated world by *«WorldMirrGpAgents»* implementation.
- Easing requirements tracing using requirement-based system structuring.

Based on the conducted validation, we observe the following limitations of our work:

- Requirement-based structuring is a roadmap for SPSys's design, increasing structural connections between system components and requirements. However, it may not be optimal for any system; therefore, we suggest further iterating the design stage with other system design methods, like those based on the Design Structure Matrix [71].

- The requirement model defines only types used in the requirement-based structuring method; however, the project team should extend the requirement model if other requirement types are necessary.
- The integrity evaluation accuracy depends on the system structure granularity, which is the resolution of the evaluation. Therefore, the more detailed the system design is, the more accurate the evaluation.
- The drawbacks pointed out in the questionnaire. Evaluation factors are objective and abstract from the subjective importance of particular components. However, such aspects can be applied by scaling the proposed factors by importance, priority, etc.

In the future, we plan to automate SPSys structure evaluation with known modelling tools like Enterprise Architect (EA). EA enables counting the components of a given stereotype. In Github,[13] we share an SQL request template. Using it in EA allows for obtaining the system components' cardinalities, which are further required for calculating the evaluation factors. However, extra effort is required to visualise the factor's values in the tool continuously. Furthermore, this work revealed the need for a CPS structure optimisation indicators taxonomy. SPSysML and future work lead to optimal structure development automation for complex robot systems like dual-arm impedance controlled mobile manipulators [12].

**CRediT authorship contribution statement**

**Wojciech Dudek:** Writing – review & editing, Writing – original draft, Visualization, Validation, Supervision, Software, Resources, Project administration, Methodology, Investigation, Funding acquisition, Formal analysis, Data curation, Conceptualization. **Narcis Miguel:** Writing – review & editing, Formal analysis. **Tomasz Winiarski:** Writing – review & editing, Writing – original draft, Visualization, Supervision, Methodology, Funding acquisition, Formal analysis, Conceptualization.

**Declaration of competing interest**

The authors declare that they have no known competing financial interests or personal relationships that could have appeared to influence the work reported in this paper.

**Acknowledgements**

The research was funded by the Centre for Priority Research Area Artificial Intelligence and Robotics of Warsaw University of Technology, Poland within the Excellence Initiative: Research University (IDUB) programme, agreement no. 1820/336/Z01/POB2/2021. The questionnaire was designed with the support of a professional poll analyst, Katarzyna Modrzejewska, to whom we pay gratitude. The work takes the robot platform and its application from the INCARE project. The authors also acknowledge TALBOT, from the European Union's Horizon 2020 research and innovation programme under Marie Skłodowska-Curie grant agreement No. 801342 (Tecniospring INDUSTRY) and the Government of Catalonia's Agency for Business Competitiveness (ACCIÓ); and SHAPES, from the European Horizon 2020 research and innovation programme under grant agreement No 857159. The Spanish grant, Spain PID2021-125535NB-I00 has also supported the work.

**Data availability**

I have shared a link to my data in the article.

---

[13] https://github.com/RCPRG-ros-pkg/spsysml

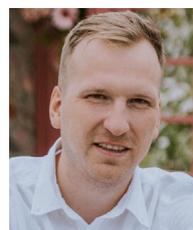

**Wojciech Dudek** IEEE, INCOSE Member, Ph.D./Eng. in control and robotics from Warsaw University of Technology (WUT); Assistant professor of WUT. Head of Safety-aware Management of robotics Interruptible Tasks in dynamic environments (SMIT) project and contributor to European Commission projects. Focused on complexity management in CPS and robot navigation, simulation and task scheduling.

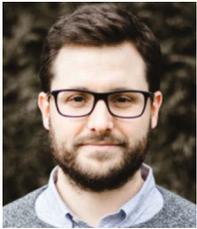 **Narcis Miguel** IEEE, Ph.D. in Mathematics from the University of Barcelona and former Postdoctoral Researcher at Politecnico di Milano. Works at Pal Robotics as Manager of the Mobile Interaction Business Unit, including product management of TIAGo family of robots and ARI.

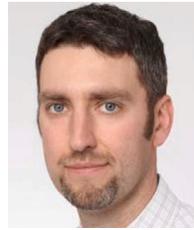 **Tomasz Winiarski** IEEE, INCOSE Member, Ph.D./Eng. in control and robotics, from Warsaw University of Technology (WUT); Assistant professor of WUT. Works on the modelling and design of robots and programming methods of robot control systems. Focused on service and social robots as well as didactic robotic platforms. Developed robotic frameworks for safe robotic research and manipulator position-force and impedance control. He recently led the WUT group in AAL – INCARE project "Integrated Solution for Innovative Elderly Care".